\documentclass[aps,prb,twocolumn,superscriptaddress,longbibliography]{revtex4-2}

\usepackage{graphicx}
\usepackage{amsmath,amssymb}
\usepackage{bm}

\begin{document}

\title{Dilute Magnetism and Edge-State Engineering in Monolayer SnO}
\author{Yuya Fukuta}
\affiliation{Department of Nanotechnology for Sustainable Energy, School of Science and Technology,
  Kwansei Gakuin University, Gakuen-Uegahara 1, Sanda 669-1330, Japan}
  \author{Souren Adhikary}
\affiliation{Department of Nanotechnology for Sustainable Energy, School of Science and Technology,
  Kwansei Gakuin University, Gakuen-Uegahara 1, Sanda 669-1330, Japan}
\author{Kazuhito Tsukagoshi}
\affiliation{Research Center for Materials Nanoarchitectonics (MANA),
National Institute for Materials Science (NIMS), Namiki 1-1, Tsukuba
305-0044, Japan}
\author{Katsunori Wakabayashi}
\affiliation{Department of Nanotechnology for Sustainable Energy, School of Science and Technology,
  Kwansei Gakuin University, Gakuen-Uegahara 1, Sanda 669-1330, Japan}
\affiliation{Research Center for Materials Nanoarchitectonics (MANA),
National Institute for Materials Science (NIMS), Namiki 1-1, Tsukuba
305-0044, Japan}
\affiliation{Center for Spintronics Research Network (CSRN), Osaka
University, Toyonaka 560-8531, Japan}
%\date{\today}

\begin{abstract}
Tin monoxide (SnO) is a p-type oxide semiconductor whose electronic properties
can be widely modified via atomic-scale engineering. Using density
 functional theory, we investigate the electronic and magnetic
 properties of transition-metal (TM = Mn, Fe, Co and W) doped SnO monolayer within a large supercell. We find that all dopants induce finite localized magnetic moments, primarily originating from $d$-orbitals of the impurity atoms. 
We show that these localized magnetic states give rise to nearly
 dispersionless bands in the vicinity of the Fermi energy (taking Co doped SnO as an example).  In addition, we investigate dimensional effects by constructing nanoribbon geometries of SnO monolayer. The ribbons exhibit intrinsic edge-localized states that are largely independent of ribbon width. For chiral nanoribbons oriented along a low-symmetry direction of the square
lattice, we find that oxygen-rich edges are thermodynamically most stable and remain semiconducting, whereas Sn-terminated edges host metallic one-dimensional
conduction channels. Our results demonstrate that transition-metal
doping and edge engineering provide effective routes to tailor the
electronic properties of SnO monolayer, making it a promising candidate for future spintronic and nanoelectronic applications.\end{abstract}

\maketitle

\section{Introduction}
Transparent and flexible electronics based on oxide semiconductors have been
extensively investigated owing to their wide band gaps, chemical stability, and
compatibility with low-temperature processing technologies, as summarized in
several comprehensive reviews
\cite{Fortunato2012OxideReview,
Kamiya2010OxideSemiconductorReview,
Hosono2006IAOSReview,
Zhang2022pTypeLonePairOxides,
Wang2016PTypeOxideReview}.
While high-performance n-type oxide semiconductors are now well established,
the realization of complementary transparent electronic circuits remains
challenging due to the limited availability of p-type oxide semiconductors with
sufficient carrier mobility and stability.

Among candidate p-type oxides, tin monoxide (SnO) is particularly attractive
because the stereochemically active Sn$^{2+}$ lone-pair states hybridize with
O $2p$ orbitals, resulting in relatively dispersive valence-band states.
Such lone-pair–driven band dispersion has been recognized as a key design
principle for achieving p-type conductivity in oxide semiconductors
\cite{Hosono2006IAOSReview}.
First-principles studies have further shown that monolayer SnO is dynamically
stable and exhibits a moderate band gap together with high intrinsic hole
mobility, making it a promising platform for two-dimensional (2D) oxide electronics
\cite{Du2017ElectronicSnOMonolayer}.
The electronic properties of SnO are further found to be highly sensitive to
atomic-scale perturbations such as strain, native defects, and surface chemistry,
which can introduce in-gap states and modify carrier transport
\cite{Shukla2020DefectsSnO_O2,Wang2022TuningSnO_OVacancy}.

Beyond intrinsic band-structure considerations, carrier transport and
stability in oxide semiconductors are known to be strongly influenced by
oxygen-related defects and dopant chemistry.
Experimental studies on In$_2$O$_3$-based amorphous oxide semiconductors
have demonstrated that controlled suppression of oxygen vacancies and
appropriate dopant incorporation are crucial for achieving stable and
high-mobility transport characteristics in thin-film transistor devices
\cite{Mitoma2014StableIn2O3TFT,Aikawa2013DopantsInOxTFT,Aikawa2015SuppressionOxygenInSiO}.
These experimental findings establish general design principles for
oxide semiconductors that are also relevant to low-dimensional SnO
systems.
Recent experiments have demonstrated that tin monoxide can
be stabilized in the atomically thin limit, retaining p-type conduction
down to thicknesses of only a few atomic layers.
This experimental realization establishes SnO as a viable 2D
oxide platform, rather than a purely theoretical model system
\cite{Huang2021SnOTFT}.

A powerful strategy to further expand the functional landscape of SnO is
transition-metal (TM) doping.
Substitutional incorporation of $3d$ TM atoms has been predicted to introduce
localized impurity states and magnetic moments, giving rise to spin-polarized
electronic structures in otherwise nonmagnetic SnO
\cite{Albar2016Magnetism3dSnO,Wang2018TMdopedSnO,Mubeen20233dTM_SnO}.
In general, TM doping can induce dilute ferromagnetic or
antiferromagnetic states in the host monolayer
\cite{hoat2024antiferromagnetism,chen2024magnetic}.
In particular, cobalt doping has been proposed as a promising route to generate
exchange-split Co-derived states near the Fermi level in SnO-based systems\cite{mubeen2023first-24a,wang2018transition-9cf}.
Most previous theoretical studies have primarily relied on density-of-states analysis. 
However, electronic transport properties are strongly influenced by band dispersion. 
Therefore, in this work, we examine transition-metal doping in SnO monolayers by analyzing the band structure in addition to the density of states.

\begin{figure*}[t]
  \centering
  \includegraphics[width=0.8\textwidth]{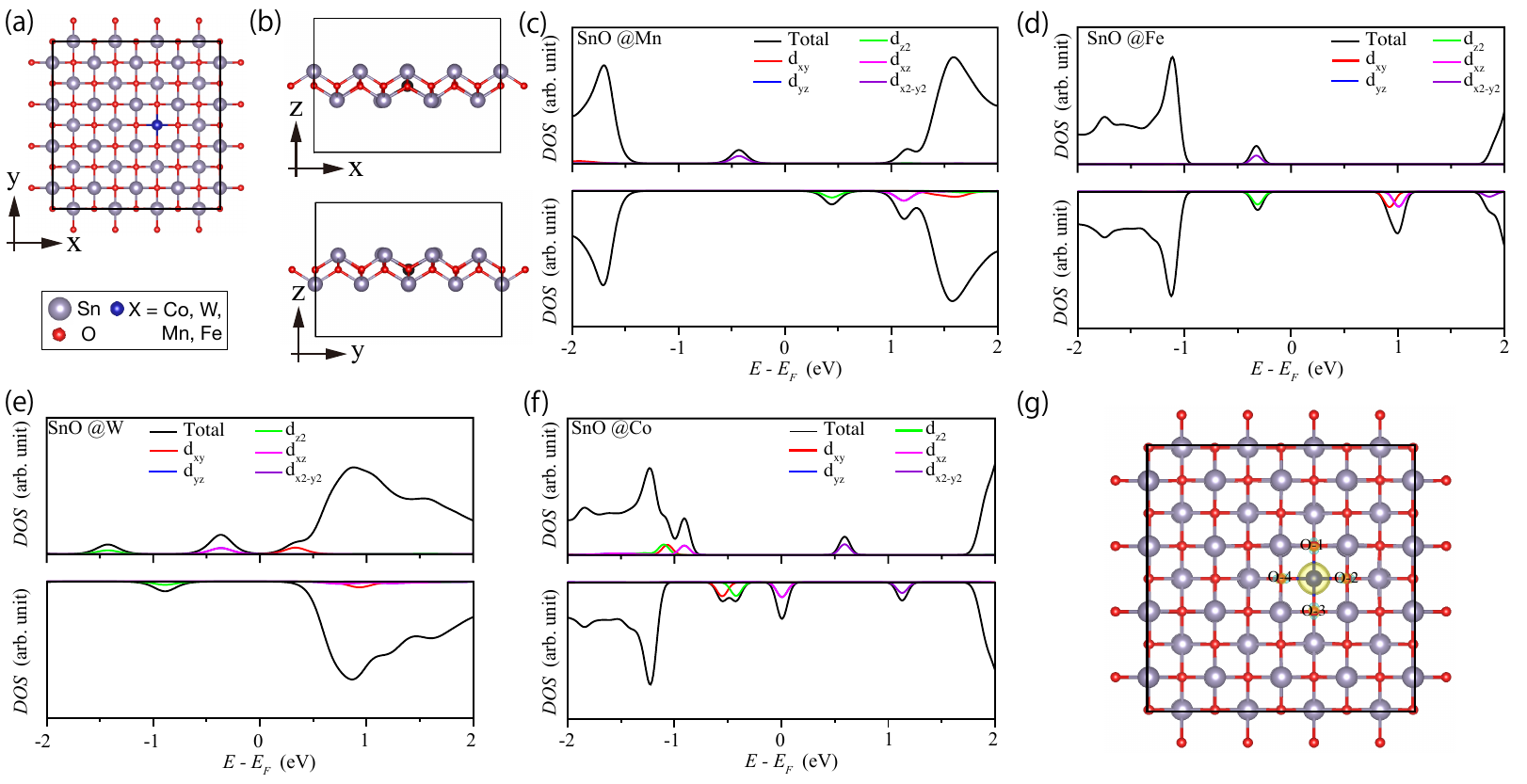}
\caption{(a), (b) Optimized top and side views of monolayer SnO structures doped with transition metals (TMs) in a $4 \times 4 \times 1$
supercell, where one Sn atom is substituted by a TM atom ( Mn, Fe, W, or Co). (c)–(f) Spin-resolved total density of states and projected density of states (PDOS) for Mn-, Fe-, W-, and
Co-doped SnO, respectively. Different color lines are showing contributions from TM $d$ orbitals. Among the considered dopants, Co doping exhibits the strongest spin asymmetry
near the Fermi level ($E_{\mathrm{F}}$), with a spin-selective electronic structure. (g) Spin-density plot of Co-doped SnO monolayer. The isosurface value is set to $0.001~e/\text{\AA}^3$.}
\label{fig:TM_doping}
\end{figure*}

Reducing SnO to finite-width nanostructures introduces an additional design
degree of freedom through edge formation\cite{wakabayashi2010edge,wakabayashi2009electronic,adhikary2024anisotropic}.
In a square-lattice system such as SnO, nanoribbons can be constructed along
high-symmetry crystallographic directions as well as along low-symmetry
(off-axis) directions with respect to the underlying lattice.
Edges oriented along low-symmetry directions provide atomic configurations that
are fundamentally distinct from high-symmetry terminations, leading to modified
local coordination environments and potentially giving rise to unconventional
edge-localized electronic states.
Despite their importance, the electronic properties of such low-symmetry edges
in SnO nanoribbons have not yet been systematically investigated.

In this work, we employ first-principles density functional theory calculations to systematically investigate the combined
effects of transition-metal doping and edge engineering in monolayer SnO. We find that all TM dopants induce finite magnetic moments, primarily originating from localized 
$d$-orbitals of the impurity atoms. In particular, substitutional Co doping introduces strongly spin-polarized mid-gap states near the Fermi level. Notably, these mid-gap states exhibit nearly dispersionless
(flat) bands in the vicinity of the Fermi level. The inclusion of on-site Coulomb interaction reveals a correlation-driven splitting of these flat bands, highlighting the importance of electron-electron interactions in
accurately describing the electronic structure. Furthermore, we show that the dispersionless bands in Co-doped SnO lead to a reduced amplitude in the optical conductivity compared to the pristine SnO monolayer.
In addition, we show that SnO nanoribbons exhibit intrinsic edge-localized states that are largely independent of ribbon width. Furthermore, nanoribbons with low-symmetry edge orientations exhibit a tunable transition between semiconducting and metallic behavior, depending on the atomic termination. These results establish atomistic design principles for controlling spin polarization, optical response, and edge-state conduction in two-dimensional SnO-based nanostructures.

\section{Computational Methods}
First-principles calculations were performed within the framework of density
functional theory (DFT) using the Vienna \emph{ab initio} Simulation Package
(VASP) \cite{Kresse1996VASP1,Kresse1996VASP2}.
The projector augmented-wave (PAW) method was employed to describe the interaction
between valence electrons and ionic cores \cite{Blochl1994PAW,Kresse1999PAW}.
Electron exchange and correlation were treated within the generalized gradient
approximation (GGA) using the Perdew-Burke-Ernzerhof (PBE) functional
\cite{Perdew1996PBE}. A plane-wave basis set with an energy cutoff of 500~eV was adopted, which was
confirmed to be sufficient to ensure convergence of total energies and electronic
structures.
Brillouin-zone (BZ) integrations were carried out using Monkhorst-Pack
$k$-point meshes \cite{Monkhorst1976MP}.
For 2D SnO monolayers, a $\Gamma$-centered mesh of $20 \times 20 \times 1$
was employed, while denser one-dimensional $k$-point sampling was used along the
periodic direction of SnO nanoribbons.
A vacuum region of at least 15~\AA{} was introduced to eliminate spurious
interactions between periodically repeated images. All atomic structures were fully relaxed until the residual Hellmann-Feynman
forces on each atom were less than 0.01~eV/\AA{} and the total energy change
between successive ionic steps was below $10^{-6}$~eV.
Spin-polarized calculations were performed for transition-metal-doped systems.
The electronic density of states and band structures were analyzed using the
\texttt{vaspkit} package \cite{Wang2021Vaspkit}. 
On-site Coulomb interaction for the Co d orbitals was included within the DFT+U framework\cite{anisimov1991band}.

To cross-check selected electronic and optical properties, additional calculations
were performed using the \textsc{Quantum ESPRESSO} (QE) package
\cite{Giannozzi2009QE,Giannozzi2017QE}.
Norm-conserving pseudopotentials were adopted as provided in the
standard QE pseudopotential libraries.

The frequency-dependent optical properties were evaluated from the complex
dielectric function $\varepsilon(\omega)=\varepsilon_1(\omega)+i\varepsilon_2(\omega)$
obtained within the linear response formalism \cite{Gajdos2006OpticsPAW,Kubo1957,Kubo1958,Greenwood1958}.
The optical conductivity $\sigma(\omega)$ was calculated using the SI-unit relation
\begin{equation}
\sigma(\omega) = \varepsilon_0 \, \omega \, \varepsilon_2(\omega),
\label{eq:conductivity}
\end{equation}
where $\omega$ is the photon angular frequency and $\varepsilon_0$ is the vacuum
permittivity.
Other optical quantities, including the refractive index $n(\omega)$, extinction
coefficient $k(\omega)$, absorption coefficient $\alpha(\omega)$, and energy-loss
function $L(\omega)$, were derived from $\varepsilon_1(\omega)$ and
$\varepsilon_2(\omega)$.

\section{Results and Discussion}
\subsection{Transition Metal Doping}
Doping with transition-metal elements is a widely explored strategy for
introducing spin polarization and magnetic functionality into semiconductors
and oxides. In particular, dilute magnetic oxides have attracted sustained interest as
potential building blocks for spintronic devices\cite{Dietl2010SpintronicsOxides}.
In the case of monolayer SnO, substitutional incorporation of $3d$ TMs such as
Co, Fe, Mn, or W provides an effective means to modify the electronic structure
by introducing localized impurity states within the band gap and inducing spin
polarization
\cite{Liang2012YdopedSnO,Tao2017TailoringSnO,Mubeen2022MnSnO,Mubeen20233dTM_SnO,Mubeen2024QuantumTransport3dSnO}.
Figs. 1(a) and 1(b) present the top and side views, respectively, of the TM-doped SnO monolayer. In this work, we consider a 4$\times$4$\times$1 supercell of the SnO monolayer, where a TM atom substitutes a Sn atom (blue-colored atom). Figs.1(c)–1(f) show the total density of states (DOS) and projected density of states (PDOS) for Mn-, Fe-, W-, and Co-doped SnO monolayers, respectively, calculated using the DFT-PBE method. Since all doped systems become spin-polarized after TM substitution, the spin-up and spin-down DOS are plotted separately.

In all cases, impurity-derived states emerge near the Fermi level and exhibit pronounced spin asymmetry. The PDOS analysis reveals that these impurity states mainly originate from the 
$d$-orbitals of the dopant atoms. For Co-doped SnO, the PDOS indicates
that the spin-up channel crosses the Fermi level, whereas the spin-down
channel remains gapped, 
suggesting an apparent half-metallic electronic structure within the
DFT-PBE
approximation\cite{Albar2016Magnetism3dSnO,Guan2017MagneticCouplingSnO}. This
behavior originates from strong hybridization between Co 3$d$ and O 2$p$
orbitals, which gives rise to spin-polarized mid-gap states. 
In Fig. 1(g), we present the spin-density distribution of the Co-doped SnO monolayer (SnO@Co). The spin density is primarily localized around the Co atom and partially distributed over the neighboring O atoms, indicating hybridization between Co 3$d$ and O 2$p$ orbitals. Table I summarizes the magnetic moments of the dopant atoms and their neighboring O atoms in the TM-doped SnO monolayers. The magnitude of the magnetic moment strongly depends on the degree of hybridization between the dopant $d$-orbitals and the neighboring O atoms.
\begin{table}[h]
\small
  \caption{Calculated local magnetic moment (in unit of $\mu$$_B$) of dopant atom and neighboring O atoms. The notations of O atoms are indicated in the Fig.1(g).}
  \label{tbl:example1}
  \begin{tabular*}{0.48\textwidth}{@{\extracolsep{\fill}}llllll}
    \hline
    System & TM atom&  O-1 & O-2 &  O-3 & O-4\\
    \hline
    SnO@Mn& 4.300 & 0.031& 0.031 & 0.031 & 0.031 \\
     SnO@Fe & 3.498 & 0.064& 0.064&0.064&0.064 \\
     SnO@W & 1.524 & 0.019&0.019&0.019&0.019 \\
     SnO@Co & 0.967 & 0.004 &0.004&0.004&0.004\\
    \hline
  \end{tabular*}
\end{table}

Thus, our PDOS analysis within the DFT-PBE approximation suggests that the Co-doped SnO monolayer exhibits half-metallic behavior. However, most previous theoretical studies have primarily focused on DOS/PDOS analysis, while the corresponding band dispersion has remained largely unexplored\cite{mubeen2023first-24a,wang2018transition-9cf,hoat2024antiferromagnetism,chen2024magnetic}. Since the localized nature of impurity-induced states plays a crucial role in determining the electronic and magnetic properties of the system, we further investigate the electronic structure through spin-resolved band structure calculations.

As a representative example, we focus on the Co-doped SnO monolayer (SnO@Co), which shows apparent half-metallicity within DFT-PBE. Fig.2(a) presents the spin-polarized electronic band structure of SnO@Co, where the red and blue bands correspond to the spin-up and spin-down channels, respectively. Notably, the bands appearing near the Fermi level are nearly dispersionless (flat) throughout the BZ. The absence of significant band dispersion indicates strong localization of the impurity-induced states around the Co atom, consistent with the spin-density distribution shown in Fig.1(g). Consequently, charge transport associated with these states is expected to be negligible due to their vanishing group velocity.

To further examine correlation effects, we performed DFT+U calculations and present the corresponding band structure for U = 3 eV in Fig.2(b) (results for other U values are provided in the Supplementary Information, see FIG.S1). The inclusion of on-site Coulomb interaction leads to a correlation-driven splitting of the flat bands near the Fermi level, resulting in the destruction of the half-metallic character predicted by standard DFT-PBE.  Note that we also varied the on-site Coulomb interaction parameter U from 1 to 2 eV (see FIG.S1). In both cases, the half-metallic character disappears; however, the overall nature of the band structure remains qualitatively similar to that obtained for U = 3 eV. These results demonstrate that the magnetic impurity states are highly localized and non-itinerant in nature, implying negligible electronic transport through these states.

\begin{figure}[t]
  \centering
  \includegraphics[width=0.45\textwidth]{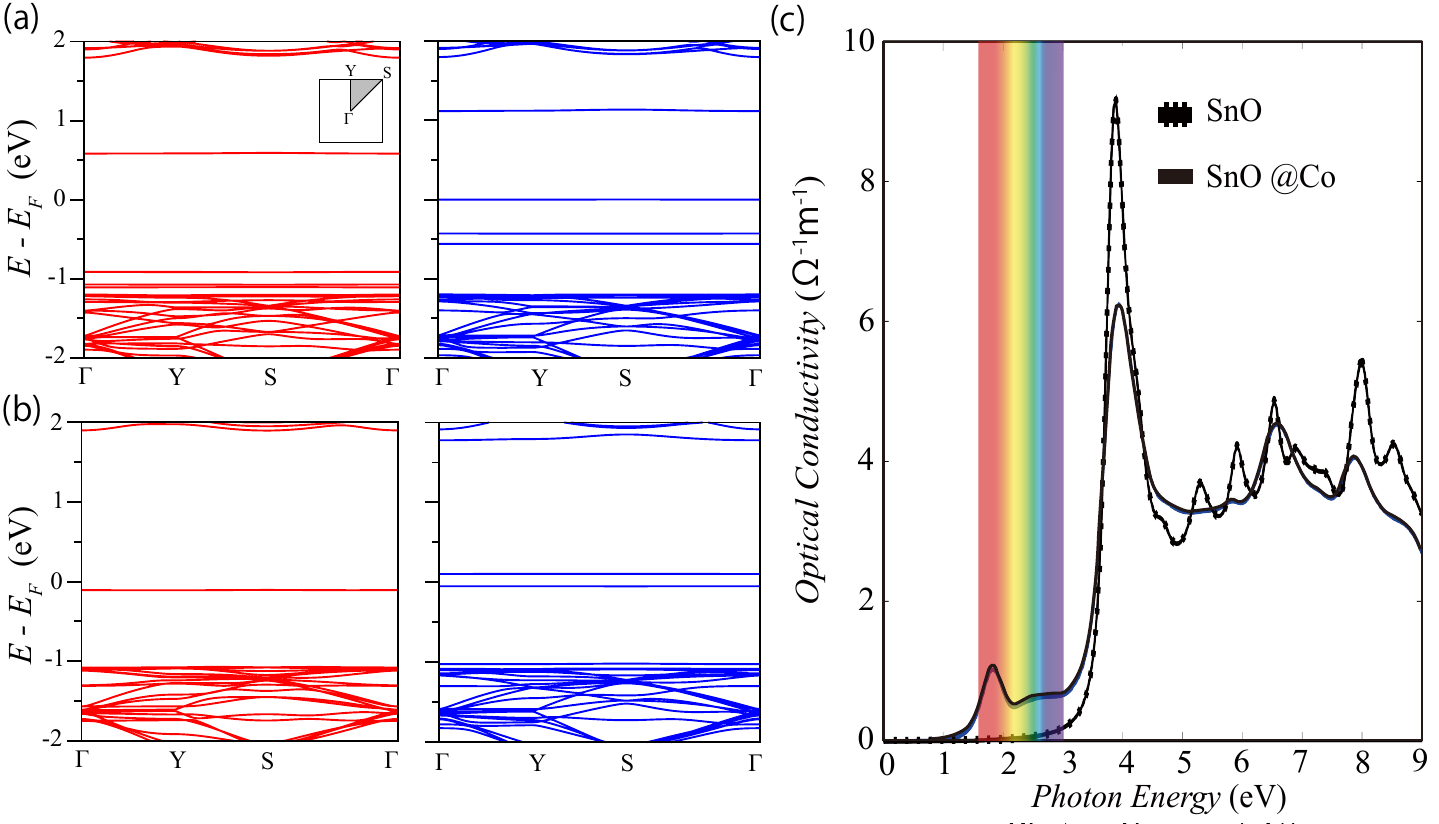}
\caption{Electronic band structure of Co doped SnO monolayer (a) without U (i.e., U = 0 eV) and (b) with U = 3 eV. Red and blue bands represent up-spin and down-spin bands respectively. The BZ is shown by the square. (c) Calculated optical conductivity spectra $\sigma$ ($\omega$) of pristine and Co-doped SnO, derived from the imaginary part of the dielectric function $\epsilon_2$ ($\omega$) according to Eq. (1). }
  \label{fig:Fig2}
\end{figure}

The non-itinerant nature of the electronic states near the Fermi level is further confirmed by the optical conductivity of the Co-doped SnO system. Fig.2(c) shows the calculated optical conductivity, $\sigma(\omega)$, for pristine and Co-doped SnO monolayers. A clear redshift of the absorption edge is observed in the Co-doped system, indicating the onset of optical absorption at lower photon energies due to the impurity-induced states near the Fermi level\cite{Fortunato2012OxideReview,Robertson2006OpticalTMoxide,Hosono2004LonePairOptical}. However, the magnitude of the optical conductivity is reduced compared to that of the pristine SnO monolayer. This suppressed optical response originates from the highly localized and nearly dispersionless nature of the Co-induced electronic states, which possess negligible carrier mobility. Overall, our results demonstrate that Co doping in SnO monolayers induces dilute magnetism accompanied by non-itinerant electronic states, highlighting the importance of localization and electron correlation in determining the electronic and optical properties of the system.

\subsection{Edge States in SnO Nanoribbons}
From a technological viewpoint, it is noteworthy that state-of-the-art
silicon devices have already achieved effective channel widths of
approximately 10~nm and are rapidly approaching the single-nanometer
scale. At such dimensions, edge and boundary effects inevitably play a central
role in determining electronic transport, rendering nanoribbon models
particularly relevant for exploring intrinsic low-dimensional physics~\cite{Auth2017IEDM10nm}.

To examine how edge structures influence the electronic properties of SnO,
we constructed hydrogen-terminated SnO nanoribbons and carried out
first-principles DFT calculations.
While edge-induced electronic states are well established in low-dimensional
lattice systems, as originally revealed in early theoretical studies and later
elaborated by first-principles calculations
\cite{Fujita1996,wassmann2008gnr},
their manifestation in square-lattice oxide nanoribbons such as SnO remains
largely unexplored. Figure 3 (a) displays the atomic geometry, electronic band structure, and wavefunction distributions for a representative SnO nanoribbon. Although hydrogen passivation is introduced primarily to eliminate dangling-bond artifacts, the edge-localized states persist even after passivation. Their energetic positions and spatial distributions are modified depending on the specific edge chemistry, indicating that these states are intrinsic features of the ribbon geometry rather than passivation-induced states\cite{wassmann2008gnr}.

%To evaluate the thermodynamic stability of different edge structures, we
%considered three passivation schemes: (i) without hydrogen termination,
%(ii) hydrogen bonded only to Sn atoms, and (iii) hydrogen bonded to both
%Sn and O atoms. 
%The corresponding relative total energies were calculated using QE to
%compare the energetic stability of different edge terminations, and are
%summarized in Table II. 
%Among the three configurations, the fully hydrogen-passivated Sn--H--O
%edge is the most stable, being 34.54 eV per supercell lower in energy
%than the unpassivated edge. The Sn--H terminated configuration is also
%stabilized by 16.26 eV per supercell. 
%These results confirm that complete hydrogen passivation is
%thermodynamically favored and provides a realistic structural model for
%investigating the intrinsic edge states of SnO nanoribbons. 

To evaluate the relative stability of different edge structures, we
considered three passivation schemes: (i) without hydrogen termination,
(ii) hydrogen bonded only to Sn atoms, and (iii) hydrogen bonded to both
Sn and O atoms. The corresponding relative total energies were calculated
using QE and are summarized in Table II. Such energetic comparisons are
commonly employed to evaluate the stability of low-dimensional
edge-terminated nanostructures \cite{PaezOrnelas2021MoSSeQD}.
Among the three configurations, the fully hydrogen-passivated Sn--H--O
edge is the most stable, being 34.54 eV per supercell lower in energy
than the unpassivated edge. The Sn--H terminated configuration is also
stabilized by 16.26 eV per supercell. These results indicate that
complete hydrogen passivation is energetically favorable and provides a
realistic structural model for investigating the intrinsic edge states
of SnO nanoribbons.

\begin{table}[t]
  \centering
  \caption{Relative total energies of hydrogen-terminated SnO nanoribbons
  with different edge configurations. Energies are given per supercell
  and referenced to the unpassivated ribbon. The values are rounded to
  two decimal places.}
  \label{tab:formation_energy}
  \begin{tabular}{c|c}
    Termination & Relative energy (eV) \\
    \hline
    Without H & 0.00 \\
    Sn--H     & -16.26 \\
    Sn--H--O  & -34.54 \\
  \end{tabular}
\end{table}

\begin{figure*}[t]
  \centering
  \includegraphics[width=0.8\textwidth]{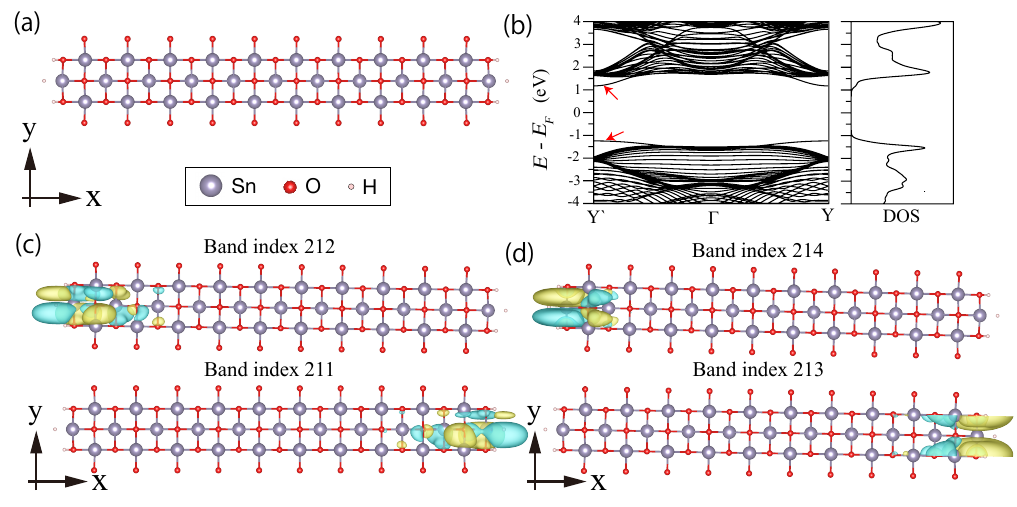}
\caption{Atomic structure of a hydrogen-terminated SnO nanoribbon, where
Sn, O, and H atoms are represented by purple, red, and white spheres,
respectively. (b) Calculated electronic band structure and density of states (DOS), showing
additional in-gap bands (shown by the red arrows) near the Fermi level ($E_{\mathrm{F}}$) that are absent
in the pristine monolayer. (c) and (d) Real-space isosurfaces of the wavefunctions corresponding to the
in-gap bands (band indices 211–212 degenerate valence bands and 213–214 are degenerate conduction bands).
The charge density is strongly localized at the ribbon edges, consistent with edge-derived electronic states.
Each in-gap band is doubly degenerate, reflecting the presence of two structurally equivalent edges in the nanoribbon.
The isosurface value is set to $0.001~e/\text{\AA}^3$.}
  \label{SnO-ribbon}
\end{figure*}

In Fig.3(b), we present band structure and DOS of SnO nanoribbon. From
the band structure additional bands (marked by the red arrow) appear in
the vicinity of Fermi 
level, which are absent in the pristine monolayer. The corresponding
wavefunction isosurfaces in Figs.3(c) and (d) demonstrate that these
states are strongly confined to the Sn- and O-terminated ribbon
edges. Each in-gap band is doubly degenerate (see band index),
reflecting the presence of two equivalent edges. Such strong
localization suggests that the edge states can function as
one-dimensional conductive channels running along the ribbon
boundaries\cite{Mitoma2014StableIn2O3TFT,Aikawa2013DopantsInOxTFT}.  
Furthermore, we confirm that these edge states are largely independent
of the ribbon width. To verify this behavior, we investigated
width-dependent electronic structures of SnO nanoribbons with both
larger and smaller widths compared to the ribbon shown in Fig.3 (a). The
corresponding results are presented in FIG.S2 of the Supplementary
Information. In all cases, edge-localized states persist, while only the
band gap changes with ribbon width. These results demonstrate that the
emergence of edge states is an intrinsic feature of the SnO nanoribbon
geometry and is robust against variations in ribbon width.

Importantly, the edge states appear for all edge termination types,
demonstrating that they are intrinsic electronic features of SnO
nanoribbons rather than artifacts of surface chemistry.
Such sensitivity to local coordination and defect chemistry is a
characteristic feature of oxide semiconductors in general.
In experimental oxide systems, including In$_2$O$_3$-based thin-film
transistors, transport properties have been shown to depend critically
on oxygen vacancy concentration and local bonding environments
\cite{Mitoma2014StableIn2O3TFT,Aikawa2015SuppressionOxygenInSiO}.
Their energetic tunability through edge passivation indicates that the
electronic structure of the ribbon boundaries can be systematically
controlled, providing a practical route for engineering
edge-dominated conduction channels in low-dimensional oxide
nanostructures\cite{Lino2011TiO2NanoribbonEdgeStates}.

\subsection{Chiral Edge Effects}
\begin{figure*}[t]
  \centering
  \includegraphics[width=0.8\textwidth]{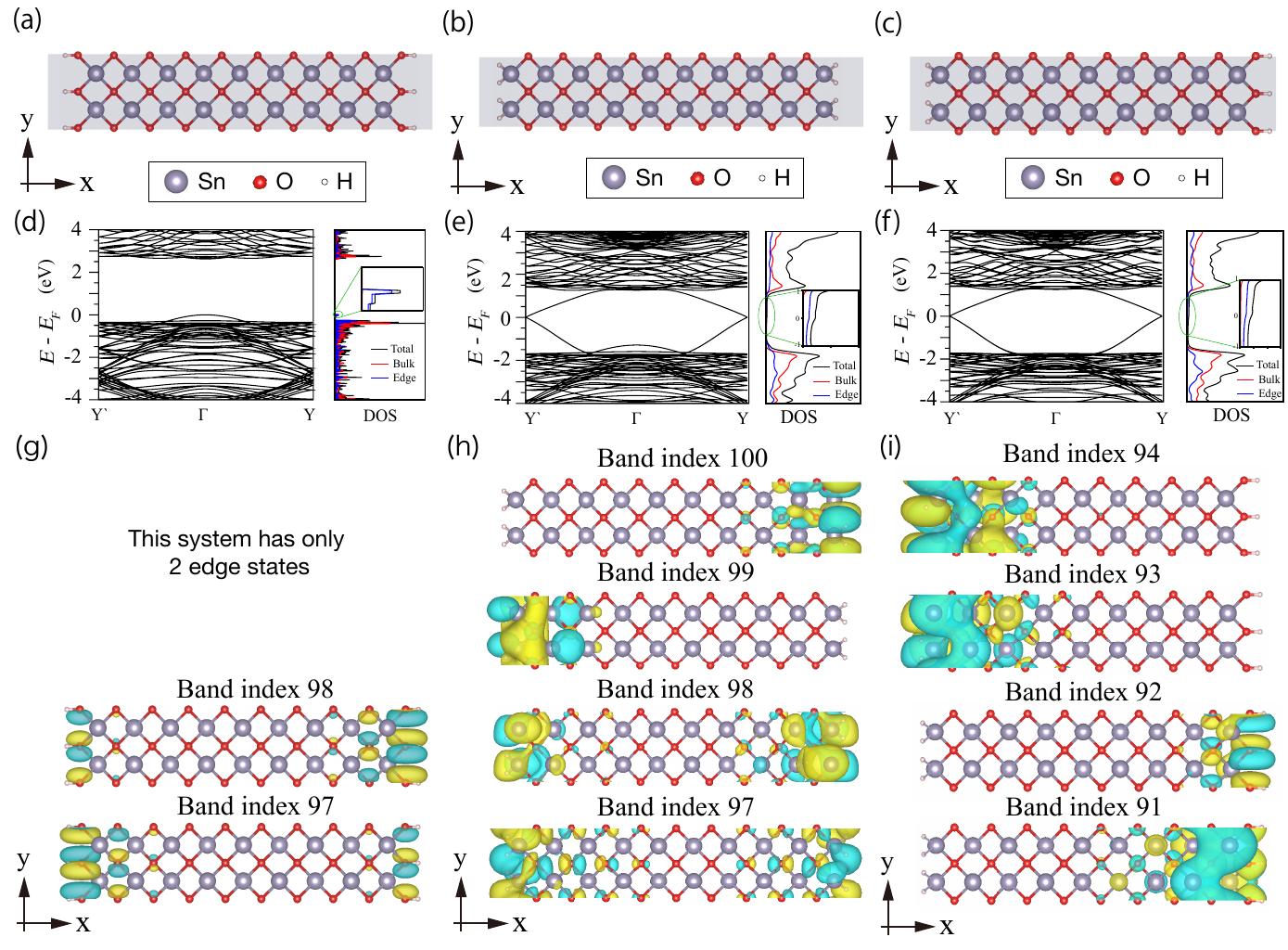}
\caption{(a)–(c) Atomic structures of hydrogen-passivated SnO nanoribbons oriented along a low-symmetry ($45^{\circ}$) direction with respect to the square lattice. The three edge terminations considered are (a) O–O, (b) Sn–Sn, and (c) Sn–O, which give rise to chemically and structurally inequivalent ribbon edges.
(d)–(f) Corresponding electronic band structures and densities of states (DOS). For the O–O terminated ribbon, edge-derived states remain localized within the band gap without crossing the Fermi level, preserving semiconducting behavior. In contrast, the Sn–Sn and Sn–O terminated ribbons exhibit metallic edge states that intersect the Fermi level, indicating the formation of one-dimensional conducting channels. In DOS, Red line and blue line is showing bulk and edge atoms contribution. The zoomed-in view shows in inset.  (g)--(i) Real-space wavefunction isosurfaces shown below each panel confirm strong localization of these states at the ribbon edges. The wavefunctions are evaluated at k-points located at the midpoint between $\Gamma$ and Y for all three ribbon configurations. The two colors represent opposite phases of the wavefunction, with yellow and cyan corresponding to positive and negative signs, respectively.}
  \label{fig:chiral_edges}
\end{figure*}

To further clarify how edge geometry influences the electronic properties of
SnO, we constructed nanoribbons oriented along a low-symmetry (45$^\circ$)
direction with respect to the underlying square lattice.
This chiral orientation corresponds to a cutting direction that is not aligned
with the principal crystallographic axes, leading to atomic edge configurations
that are not related by mirror or rotational symmetry.
As a result, the two ribbon edges become chemically and structurally inequivalent,
giving rise to distinct electronic environments.
As shown in Figs.~\ref{fig:chiral_edges}(a)–(c), three representative edge
terminations were examined: (i) oxygen-terminated edges (O–O),
(ii) tin-terminated edges (Sn–Sn), and (iii) mixed tin–oxygen termination (Sn–O).
All structures were hydrogen-passivated using the most stable scheme identified
previously.

DFT calculations reveal that all three chiral SnO nanoribbons host
edge-derived electronic states within the fundamental band gap.
The nature of these states depends strongly on the edge composition:
the O--O terminated ribbon [Fig.~\ref{fig:chiral_edges}(d)] remains
semiconducting, whereas the Sn--Sn and Sn--O terminations
[Figs.~\ref{fig:chiral_edges}(e) and (f)] exhibit metallic behaviour,
with edge states intersecting the Fermi level. We analyze the contributions of edge and bulk atoms through PDOS calculations, where the blue and red curves represent the edge and bulk atom contributions, respectively. For all three edge configurations, the electronic states near the Fermi level are found to originate predominantly from the edge atoms, as clearly shown in the zoomed-in views of the DOS plots. This confirms the edge-localized nature of the states appearing around the Fermi energy. Such termination-dependent emergence of edge-localized bands is a
general feature of oxide nanoribbons, in which reduced coordination and
edge chemistry govern both electronic structure and stability
\cite{Lino2011TiO2NanoribbonEdgeStates,Topsakal2009ZnOHoneycomb,BotelloMendez2008ZnONanoribbon}.

Wavefunction isosurfaces plotted [see Fig.~\ref{fig:chiral_edges}(g)--(i)] beneath each band structure confirm the
strong spatial localization of these states at the ribbon boundaries.
While the O--O termination yields localized in-gap states that do not
contribute to conduction, Sn-containing edges support boundary states
that extend along the ribbon direction, consistent with one-dimensional
metallic channels. Furthermore, we plot the charge density distributions for the three types of edge nanoribbons in FIG.S3. The charge density analysis reveals that the Sn-terminated edge forms a delocalized charge channel along the Sn atoms, which gives rise to the metallic nature of the Sn-edge SnO nanoribbon.

To assess the relative stability of the three configurations,
relative energies were computed.
The relative energies of the O–O, Sn–Sn, and Sn–O
terminated ribbons are -5.48 eV, -4.92 eV, and
-5.17 eV per unit cell, respectively.
The results 
show that the O–O terminated ribbon has the lowest formation energy,
making it the most stable among the three. 
Although the Sn–Sn and Sn–O terminations exhibit metallic
edge conduction, the O-rich configuration is energetically
more stable.

The enhanced stability of the O–O terminated edge can be attributed to the
effective saturation of dangling bonds and the resulting suppression of
localized high-energy states at the boundary.
Such a strong correlation between defect passivation, local coordination,
and electronic transport is a recurring theme in oxide semiconductors,
where the balance between structural stability and carrier delocalization
critically governs device performance
\cite{Wang2016PTypeOxideReview,Huang2021SnOTFT}.
In contrast, Sn-rich edges inherently retain partially filled Sn-derived
orbitals, which promote metallic edge conduction but incur a higher energetic
cost.
This trade-off between structural stability and electronic conductivity
highlights a general design principle for square-lattice oxide nanoribbons:
oxygen-rich terminations favor thermodynamic stability, whereas cation-rich
edges enable low-dimensional metallic transport.

These results reveal a clear trade-off between structural stability and
electronic conductivity in SnO nanoribbons.
Oxygen-rich edge terminations effectively saturate dangling bonds and
thus minimize high-energy localized states, leading to enhanced
thermodynamic stability.
In contrast, tin-rich edges inherently retain partially filled
Sn-derived orbitals, which promote metallic edge conduction at the cost
of higher formation energies.
Such a balance between energetic stability and low-dimensional
conductivity reflects a general design principle in low-dimensional
oxide systems, where local coordination and defect chemistry play
decisive roles in determining electronic functionality
\cite{Robertson2021DopingLimitsPtypeOxides}.
Accordingly, chiral edge engineering combined with controlled edge
passivation provides a viable route for tailoring the electronic
properties of SnO nanoribbons.

\section{Conclusion}
In conclusion, we have systematically investigated the electronic,
magnetic, and optical properties of monolayer SnO through transition-metal
doping and edge engineering using first-principles calculations.
Substitutional Co doping introduces strongly spin-polarized impurity
states near the Fermi level, 
giving rise to an apparent half-metallic behavior within the DFT-PBE
approximation. 
However, upon inclusion of on-site Coulomb interaction, 
the half-metallic character disappears due to correlation-driven splitting of the localized impurity states. These findings indicate that Co-doped SnO exhibits dilute localized magnetism, highlighting its potential for oxide-based spintronic applications.

We further demonstrated that SnO nanoribbons host robust edge-localized
electronic states that persist across different edge terminations and
passivation schemes, confirming their intrinsic origin. In addition, these localized edge states are largely independent of the ribbon width.
Hydrogen passivation stabilizes the edge structures without eliminating
the edge states, providing a realistic platform for probing
edge-dominated transport in oxide nanostructures.
For chiral nanoribbons oriented along low-symmetry directions, a clear
trade-off emerges between thermodynamic stability and electronic
conductivity: oxygen-rich edges favor structural stability, whereas
tin-rich edges enable metallic one-dimensional conduction channels.

The present results also suggest that the proposed strategy of combining transition-metal doping and edge engineering may be extended to other oxide semiconductors. However, the emergence of localized flat-band states and edge-dependent metallicity in SnO is strongly influenced by its unique lone-pair-driven electronic structure arising from Sn 5s-O 2p hybridization, which distinguishes it from oxide systems such as SnO$_2$, In$_2$O$_3$. Therefore, while the general design principles may be broadly applicable, the resulting electronic and magnetic behaviors are expected to depend sensitively on the underlying orbital characteristics of the host oxide material.

These findings establish general design principles for low-dimensional
oxide semiconductors, where electronic functionality can be tuned through
a delicate interplay of dopant chemistry, local coordination, and edge
composition.
The present work thus positions monolayer SnO as a versatile platform
for exploring multifunctional nanoelectronic and spintronic phenomena in
oxide-based two-dimensional materials.

\section*{Acknowledgements}
This work was supported by JSPS KAKENHI (Grants No. JP25K01609,
No. JP22H05473, and No. JP21H01019), JST CREST (Grant
No. JPMJCR19T1). K.W. acknowledges the financial support for Basic
Science Research Projects (Grant No. 2401203) from the Sumitomo
Foundation. 

\section*{DATA AVAILABILITY}
All relevant data supporting the findings of this study are included in
the article and its Electronic Supplementary Information
(ESI). Additional data or computational files are available from the
corresponding author upon reasonable request.

\bibliographystyle{apsrev4-2}
\bibliography{references-2}

\end{document}